\input{psfig.sty}
\documentclass[preprint]{aastex}

\begin{document}
 
\title{The 0.1-100~keV Spectrum of LMC~X--4 in the High State:
Evidence for a High Energy Cyclotron Absorption Line}

\author{A. La Barbera\altaffilmark{1}, 
L. Burderi\altaffilmark{2},
T. Di Salvo\altaffilmark{1},
R. Iaria\altaffilmark{1},
N. R. Robba\altaffilmark{1}}
\altaffiltext{1}
{Dipartimento di Scienze Fisiche ed Astronomiche, Universit\`a di Palermo, 
via Archirafi 36 - 90123 Palermo, Italy}
\authoremail{nino@gifco.fisica.unipa.it}
\altaffiltext{2}{Osservatorio Astronomico di Roma, Sede di Monteporzio
Catone - via di Frascati 33 - 00044 Roma, Italy}

\begin{abstract}

We report the spectral analysis of the X-ray pulsar LMC~X--4 
in its high state out of eclipse observed by BeppoSAX.
During this observation no coherent pulsations are detected. 
The primary continuum is well described
by a power law with a high energy cutoff 
($E_{\rm cutoff} \sim E_{\rm fold} \sim 18$~keV).
The addition of a cyclotron
absorption line at $\sim 100$~keV improves the fit significantly.
The inferred magnetic moment is $1.1 \times 10^{31}$ Gauss cm$^3$, 
in agreement with the value estimated assuming that the neutron star is
at the spin equilibrium, as it has been proposed for this source. 
The remaining excess at low energies can be fitted by
a Comptonization of soft photons by moderately hot electrons 
(kT $\sim 0.9$~keV), with an optical depth $\tau \sim 16$. 
The seed photons for this Comptonization are consistent with black body 
emission from the accretion disk at the magnetospheric radius.
Another possibility is to fit the soft excess with black body and thermal
bremsstrahlung. In this case the black body would originate from cold
plasma at the magnetosphere while the bremsstrahlung component may
be produced by the strong stellar wind from the companion star, ionized
by the X-ray emission from the pulsar.

\end{abstract}   


\section{Introduction}

LMC~X--4 is an eclipsing high-mass X-ray binary pulsar
in the Large Magellanic Cloud. 
Its optical counterpart was identified with a 20 $M_{\odot}$ O7 III-V
star (Sandaleak \& Philip 1977). 
The orbital period of the system is 1.4 days
(Li, Rappaport \& Epstein 1978; White 1978). 
The X-ray intensity varies
by a factor $\sim$ 60 between high and low states with a cycle 
of 30.3 days (Lang {\it  et al.} 1981). 
This long-term variation is attributed to a periodic blockage
of the direct X-ray beam by a precessing accretion disk, which is tilted
with respect to the orbital plane of the binary
(Lang {\it  et al.} 1981; Ilovaisky et al 1984, Priedhorsky \& Holt 1987;
Woo, Clark \& Levine 1995, and references therein). 
Flaring episodes occur about 
once per day, lasting from $\sim$ 20 s to 45 minutes,
during which the intensity increases by factors up to 
$\sim$ 20 (Epstein {\it  et al.} 1977; White 1978; Skinner {\it  et al.} 
1980; Kelley {\it  et al.} 1983; 
Pietsch et al 1985; Dennerl 1989; Levine {\it  et al.} 1991).

Coherent pulsations with a period of 13.5 s were discovered by 
Kelley {\it  et al.} (1983) during flares events. 
Later, the 13.5 s periodic 
pulsations were also detected in EXOSAT observations
during non-flaring out-of-eclipse state (Pietsch {\it  et al.} 1985).
Nevertheless, while the periodic pulsations have been revealed several times
during flaring events
(Levine {\it  et al.} 1991, Woo, Clark \& Levine 1995, Woo {\it  et al.} 1996), only
once periodic pulsations have been observed during non-flaring state.

The orbital period decreasing rate is 
$\dot{P}_{\rm orb}/P_{\rm orb} = (-5.3 \pm 2.7) \times 10^{-7}$ yr$^{-1}$.
This value is smaller than that measured for
Cen X--3 and SMC X--1 placing a lower limit 
on the rate of orbital energy dissipation by tidal forces. 
This fact, together with the relatively large radius of the companion
(14 $\leq$ R $\leq$ 20 $R_{\odot}$), suggests
that the companion is expanding rapidly, still burning hydrogen
into a shell (Levine {\it  et al.} 1993).
In fact the resulting increase of the moment of inertia could 
acts to reduce the rotation rate of the companion star producing a 
relatively large difference between stellar rotation period and orbital
period. This enhances the tidal friction effects producing a large orbital
decay.

Pulse-phase resolved spectroscopy from Ginga and ROSAT 
observations in the energy range 0.2--30~keV 
shows that the phase dependent spectrum can be 
modeled as the sum of three component (Woo {\it  et al.} 1996): 
a high energy component represented
by a power law with photon index $\alpha \sim 0.7$ with 
a high energy cutoff at 16.1~keV and a folding energy of 35.6~keV,  
a thermal bremsstrahlung component with temperature of 0.35~keV
and a pulse-phase independent black body (kT $\sim$ 0.03~keV). 
An iron line is also present at E $\sim 6.6$~keV.

Several observations of the spin period evolution
showed both spin up and spin down episodes.
This suggests that the pulsar has reached a spin equilibrium 
state, wherein its spin period approximately equals the Keplerian period at 
the inner edge of the accretion disk. Because of the high X-ray luminosity
of LMC~X--4 and the relatively slow spin period, a very strong magnetic 
dipole has to be present ($\geq 10^{31}$ G cm$^3$) (Woo {\it  et al.} 1996; 
Naranan {\it  et al.} 1985).

Up to date no clear evidence of a cyclotron absorption
feature has been reported. This is in line with the fact that the
value of the cyclotron line energy deduced from the 
spin equilibrium ($\sim$ 100~keV) 
is out of the energy range (typically up to 30~keV)
of most X-ray satellites which antedate BeppoSAX.
In fact the centroid of the line is related to the magnetic moment
by the relation $E_{\rm cycl} = 11.6 \mu_{30}/R_6^3 (1+z)$ 
where $\mu_{30}$ is the magnetic dipole moment in units of $10^{30}$ G cm$^3$,
$R_6$ is the neutron star radius in units of  $10^{6}$ cm,
and $(1+z) = (1 - 2G M_{\rm NS}/R c^2)^{-1/2}$ 
is the gravitational redshift factor. 
Adopting a standard neutron star mass of 1.4 
$M_{\odot}$ with a standard radius of $10^{6}$ cm and $\mu \geq 10^{31}$ G 
cm$^3$, we obtain $E_{\rm cycl} \geq 90$~keV.
The broad spectral band of the BeppoSAX satellite, extending up to
200~keV, gives us the possibility of studying the spectrum up to the 
energies where the cyclotron absorption feature is expected.

In this paper we report the spectral analysis of LMC~X--4 out of eclipse
in high state observed by BeppoSAX in the 0.12--100~keV
energy range. We confirm the complex multicomponent 
nature of the spectrum of LMC~X--4. Moreover we find that there are
evidences of  the presence of a cyclotron line, with the centroid 
around 100~keV, at a high degree of confidence.
  
\section{Observations and Light Curves}
BeppoSAX observed LMC~X--4 with its narrow-field instruments 
(NFI; Boella {\it  et al.} 1997a) in 1998 from October 20th
22:40 to October 22th 08:05 (UT) corresponding 
to the phase interval 0.66--0.70 of the 30.3 days cycle, 
when the source was in its high state. 
The ephemeris (period: 30.3 $\pm$ 0.05 days, epoch: 50087 MJD, corresponding 
to the beginning of the low-state phase) has
been obtained by using the observations of LMC~X--4 performed by the
ASM on board RXTE.

The BeppoSAX observatory covers more than three decades of energy,
from 0.1 to 200~keV. The payload is composed by four coaligned instruments:
the Low-Energy Concentrator Spectrometer (LECS, 0.1--10~keV; 
Parmar {\it  et al.} 1997), 
the Medium-Energy Concentrator Spectrometer consisting of two units
(MECS, 1--10~keV; Boella {\it  et al.} 1997b), 
the High-Pressure Gas Scintillation Proportional Counter 
(HPGSPC, 4--120~keV; Manzo {\it  et al.} 1997), and the Phoswich
Detection System (PDS, 15--300~keV; Frontera {\it  et al.} 1997). 
The fields of view (FOV) of the instruments is approximately $1^{\circ}$. 
Since LECS and MECS have imaging capabilities, we extracted the data from 
circular regions in the LECS and MECS FOVs of 8' and 4' radius, respectively, 
centered on the maximum of the point-spread function (PSF). 
The background subtraction is obtained using 
blank-sky observations and extracting the background spectra from a
circular region corresponding to that one used for the source. The HPGSPC and
PDS do not have imaging capabilities. In this case the subtraction of the 
background is obtained using off-source data collected during the rocking 
of the collimators. The effective exposure during the observation 
are so summarized: 
$\sim$ 28 ks for the LECS, $\sim$ 66 ks for the MECS, $\sim$ 29 ks 
for the HPGSPC, and $\sim$ 39 ks for the PDS.

In Figure 1 (upper panel) we show the light curve of LMC~X--4, binned at 
300~s, in the energy band 1.5--10.5~keV (MECS data).
The light curve shows that the  X-ray source is eclipsed
for about 20 ks. No important flaring episodes are observed in 
this light curve. The corresponding hardness ratio, i.e. the ratio
of the counts in the 4.5--10.5~keV energy band to the counts in the
1.5--4.5~keV energy band, is also shown in Figure~1 (lower panel).
Although the source flux increases by $\sim 20-30$\% during the 
pre-eclipse phase, 
the hardness ratio appears to be quite constant during our observation.
Actually the hardness ratio decreases from $\sim 2.5$ to $\sim 1$ in a
short time interval just after the eclipse.  In the following analysis 
we considered only data out of eclipse, i.e.
we excluded the time interval from $8.7 \time 10^4$~s to 
$1.1 \times 10^5$~s, containing both the eclipse and the short interval
where the hardness ratio varies significantly.

The average flux out of eclipse in the high state is 
$9.7 \times 10^{-10}$ ergs cm$^{-2}$ s$^{-1}$ in the 0.1--100~keV band. 
Adopting a distance to the source of 50 kpc, the resulting luminosity is 
$\sim 3 \times 10^{38}$ ergs s$^{-1}$. 
In this paper we concentrate our analysis on data out of eclipse.

Because of the better statistics, we used MECS data to investigate for 
periodicities.
We searched for periodic pulsations in a large interval (12.5-14.5 s)
around the value of the spin period reported in the literature, 13.5 s,
by using folding techniques. No pulsations have been found.
In fact the maximum excess that we have found had $\chi^2 = 93.82$ 
for 63 degrees of freedom, which is not significant.

\section{Spectral Analysis}

Spectral analysis was performed on the average photon spectrum of 
LMC~X--4, out of eclipse, in the energy range 0.12--100~keV. 
The energy ranges used for each NFI are: 0.12--3~keV for LECS, 
1.8--10~keV for MECS, 9--30~keV for HPGSPC, and 
15--100~keV for PDS. Different normalizing factors for the four 
instruments were included, fixed to 1 for the MECS and kept 
free for the other instruments.

The continuum is described by a power law with a high energy cutoff
corrected by photo-electric absorption.
Indeed, by using the XSPEC models, POWERLAW and HIGHECUT, an evident residual
remains at the cutoff energy. Since it just appears at the cutoff energy, 
where the fitting function shows a cusp,
this residual probably is an artifact of the model.
To verify this hypothesis we changed the XSPEC HIGHECUT
function, whose mathematical form is:
\begin{equation}
H(E) = \left\{ \begin{array}{ll}
 1              & \mbox{for $E \leq E_{\rm cutoff}$}\\
 exp(\frac{E_{\rm cutoff} - E}{E_{\rm folding}}) & \mbox{for $E > E_{\rm cutoff}$}
       \end{array}
       \right .
\end{equation}
with a function $F(E)$ that avoids the cusp. In fact
$F(E)$ has a smoothed region of width W, between the low energy range 
($E \leq E_{\rm cutoff} - W/2$)
and the high energy range ($E \geq E_{\rm cutoff} + W/2$)
described by a third degree polynomial function, in order to obtain
a function that is continuous with its derivatives
(see also Burderi {\it  et al.} 2000);
The mathematical form of our function is:
\begin{equation}
F(E) = \left\{ \begin{array}{ll}
 1              & \mbox{for $E \leq E_{\rm cutoff} - W/2$}\\
 AE^3+BE^2+CE+D & \mbox{for $E_{\rm cutoff} - W/2< E < E_{\rm cutoff} + W/2$}\\
 exp(\frac{E_{\rm cutoff} - E}{E_{\rm folding}}) & \mbox{for $E \geq E_{\rm cutoff} + W/2$}
       \end{array}
       \right .
\end{equation}
where A, B, C, D are constants 
calculated imposing continuity conditions for the function F(E) 
and its derivatives while W is left as a free parameter of the fit.
Using this model, the residuals at the cutoff energy disappear, and
the fit improves significantly.

An evident soft excess remains at low energies below $\sim$ 2~keV. 
A simple black body model is not a good description of the soft excess,
giving $\chi^2/d.o.f. \sim 727/537$. 
Following the literature (Woo {\it  et al.} 1996) we fit this feature
with a thermal bremsstrahlung emission plus a black body.
The resulting parameters describing the model are reported in Table 1 
(model 3 and 4) and can be so summarized. 
The photon index of the power law is $\sim 0.6$, the 
cutoff energy is $\sim 18$~keV and the folding energy is
$\sim 17.5$~keV. The photo-electric absorption corresponds to
an hydrogen column of $\sim 5 \times 10^{20}$ cm$^{-2}$. The temperature
associated to the bremsstrahlung is 0.85~keV while that of the black body
is $\sim 6 \times 10^{-2}$~keV. 

An alternative way of fitting the soft excess is to model it
with a Comptonization model (see Table 1, model 1 and 2) . 
We used the COMPST model of XSPEC (v.10). 
We preferred COMPST to COMPTT because
COMPST needs less free parameters. In addition, using COMPTT, the 
temperature of the seed photons obtained by the fit is less than 0.1~keV,
i.e. it is out of the spectral range, making this estimate unreliable.
The temperature associated with
the hot electrons is $\sim 1$~keV and the relative optical depth is
$\sim 16$. The parameters of the other model components remain 
substantially unchanged. 

The presence of a gaussian emission line at $\sim$ 
6.5~keV is evident. Indeed the $\chi^2$
reduces significantly if we model it with two gaussians centered at the 
energies 6.1 and 6.6~keV 
(equivalent width $\sim$ 200 and 65 eV, respectively). Comparing the model
with one gaussian and that with two gaussians,
the F-test gives a probability greater than 0.997 that the improvement is
not casual. The simultaneous presence
of a narrow line at 6.6~keV and a broader one at 6.1~keV  
implies two different iron ionization stages.
In addition an absorption edge is observed at
about 1.25~keV with a very large statistical evidence
(the F-test probability of chance improvement is less than $1 \times 10^{-6}$). 
This feature could be attributed to the absorption edge of Fe XVII (1266 eV).

Large residuals also appear in the hard part of the LMC~X--4 spectrum 
starting from 40~keV and extending up to 100~keV (see figure 2, upper panel). 
We interpreted this feature as a cyclotron absorption line and modeled it 
with a gaussian profile, i.e. we used a multiplicative function given by:
$f(E) = 1-A_{\rm cycl}\ exp[-(E-E_{\rm cycl})^2/(2\sigma^2)]$, with the 
condition that $f(E) = 0$ when it assumes negative values. 
Since the width of cyclotron lines is most probably of thermal
origin (see e.g. Mihara 1995; Dal Fiume et al. 1999; Burderi et al. 2000), 
we preferred to use a gaussian profile, instead of a lorentzian profile, 
to fit this feature. 
Using this model the residuals disappear (see Figure 2, lower panel).
The centroid energy of the line is
$\sim 100$~keV and the sigma is  $\sim 50$~keV. 
The addition of this component reduces the $\chi^2$ significantly, 
as it is evident comparing models 1 and 3 with respect to models 2 and 4
in Table 1.
The F-test gives a probability of $4 \times 10^{-4}$ for a chance 
improvement of the fit. 

Summarizing the whole 0.1--100~keV spectrum of LMC~X--4 can be described 
as follows:
\begin{equation}
S(E)=WABS \times EDGE \times CYCL \times (SOFT 
        + 2\ GAUSS + F(E) \times POW)
\end{equation} 
where WABS is the photo-electric absorption, EDGE is the absorption edge,
CYCL is the cyclotron absorption feature modeled as a gaussian, 
SOFT is the component that can be modeled as a 
bremsstrahlung plus a black body or as a Comptonization,  
2\ GAUSS represents the two
iron emission features at about 6.5~keV, F(E) is the cutoff function above 
described and finally POW is the power law component. 
The fit corresponding
to the model with bremsstrahlung and relative residuals are plotted 
in Figure 3, 
while the fit corresponding to the model with Comptonization and 
relative residuals
are plotted in Figure 5. In both cases the cyclotron absorption line 
is present. The unfolded spectra corresponding to the two scenarios are drawn 
in Figure 4 and 6, respectively.

\section{Discussion}

We have found evidence, with a high degree of confidence, 
of a high energy cyclotron
absorption feature in the broad band (0.1-100~keV) spectrum of LMC~X--4.
Modeling this feature as a gaussian in absorption,  
we have found the centroid energy 
at $\sim 100$~keV with sigma $\sim 50$~keV,
corresponding to a magnetic moment of $1.1 \times 10^{31}$ Gauss cm$^3$.
The addition of this new component 
to our model significantly reduced the $\chi^2$,
with a probability of chance improvement of $4 \times 10^{-4}$.  
This is in line with the expectation that LMC~X--4 should have a very strong 
magnetic field.  

There are some evidences that LMC~X--4 have
reached its spin equilibrium wherein its spin period is close to the 
Keplerian period at the inner edge of the accretion disk, which is probably  
the magnetospheric radius of the system. In fact several observations show 
both spin-up and spin-down regimes (see e.g. Woo {\it  et al.} 1996). 
The average values of $\dot{P}_{\rm pulse}$ range from 
$-4.0 \times 10^{-3}$ s/yr between the HEAO 2 and the first of the EXOSAT 
observations to $+1.9 \times 10^{-3}$ s/yr between EXOSAT and the most recent 
Ginga and Rosat observations. The radius $R_{\rm eq}$ at which the Keplerian 
period of the orbiting matter in the disk equals the pulsation period 
$P_{\rm pulse}$ is simply 
$R_{\rm eq} \sim 1.5 \times 10^8 m^{1/3} P_{\rm pulse}^{2/3}$ cm, 
where  $m$ is the neutron star mass in units of 
solar masses and $P_{\rm pulse}$ is the spin period in seconds. 
For $P_{\rm pulse} = 13.5$ s, $R_{\rm eq}$ results $\sim 9 \times 10^8$ cm.
In addition we know that, for an accreting magnetized neutron star, the 
magnetosphere radius, $R_{\rm m}$, is (see, e.g., Burderi {\it  et al.} 1998):
\begin{equation}
R_{\rm m} \sim 4.3 \times 10^8 \phi \mu_{30}^{4/7} L_{37}^{-2/7}
        m^{1/7} R_{6}^{-2/7} \; {\rm cm}
\end{equation}
where $\mu_{30}$ is the magnetic moment in units of $10^{30}$
Gauss cm$^3$, $L_{37}$ is the intrinsic luminosity in units of $10^{37}$ erg/s,
$R_{6}$ is the neutron star radius in units of $10^{6}$ cm,
and finally $\phi$ ($\leq 1$) is the correction to the Alfv\'en radius for
the presence of the accretion disk. Equaling  $R_{\rm eq}$ and $R_{\rm m}$ 
it is possible to estimate the strength of the magnetic moment.
Taking into account the intrinsic high luminosity of LMC~X--4 
($\sim 3 \times 10^{38}$ ergs/s in the
0.1-100~keV band), we derive a high 
magnetic dipole moment, of the order of $2 \times 10^{31}$ 
Gauss cm$^3$. If this is correct, any cyclotron spectral feature 
in this source has to be expected at energies
around 150~keV. Indeed we find evidences of an absorption feature 
in the hardest part of spectrum of LMC~X--4 centered at 100~keV which
is consistent with the expected value.

The width (gaussian $\sigma$) of the cyclotron line that we found 
is $\sim 57$~keV. If the broadening is caused by thermal Doppler effects, 
the relationship among the folding energy of the exponential cutoff, 
$E_{\rm fold}$,
that is a measure of the temperature close to the neutron star,  
the centroid energy of the cyclotron line, $E_{\rm cycl}$, and its width, 
$\sigma$, is: $\sigma \approx E_{\rm cycl} (E_{\rm fold}/m_e c^2)^{1/2}$. 
Adopting this relationship the expected width of the line would be 
$\sim 26$~keV that is almost half the measured broadening of the line.
However, note that the width of the cyclotron line might be not well 
constrained since the centroid energy of the line is at the end of the 
energy range of PDS, and therefore only the left wing of the line is visible.

The whole 0.12-100~keV spectrum of LMC~X--4 is complex.
The primary continuum is well fitted by a power law with a high energy cutoff,
smoothed at the cutoff energy. This component could be produced 
by Comptonization of the radiation coming from the hot spots on the 
neutron star by the plasma in the accretion column, very close to the neutron 
star surface.
Anyway, at low energies, an evident soft excess remains. We fitted
this excess with two different models: 1) bremsstrahlung emission 
with the addition of a black body, 2) Comptonization.

In the first scenario, 
the black body component is probably emitted by the accretion disk, whereas
the bremsstrahlung component comes from an optically thin plasma surrounding
the system.
In fact, considering that the luminosity and the temperature of the 
black body component, obtained by the fit, are 
$1.3 \times 10^{37}$ ergs s$^{-1}$ and $6.3 \times 10^{-2}$~keV,
respectively, the typical radius associated with emitting region of 
the black body component is $\sim 2.5 \times 10^8$ cm. 
This value is in agreement with that found for the
magnetospheric radius under the spin equilibrium hypothesis.
With regard to the bremsstrahlung component, the calculated luminosity is 
$2.2 \times 10^{37}$ ergs s$^{-1}$. This give us the possibility
to estimate the radius of the bremsstrahlung emitting region. 
The luminosity of the bremsstrahlung component, $L_{\rm brems}$, can be 
expressed as:
\begin{equation}
L_{\rm brems} \sim 5.8 \times 10^{-24} T^{1/2} N_e^2 V \; {\rm erg/s}
\end{equation}
where $T$ is the electron temperature in~keV, $N_e$ is the electron density
and V is the volume of the emission region in cm$^3$. Considering the 
relationship between the optical depth $\tau$ and $N_e$: 
$\tau \sim \sigma_{\rm T} N_e R$, where $\sigma_{\rm T}$ is the Thomson cross 
section and $R$ is the radius of the emission region, and supposing a spherical 
geometry, we obtain:
\begin{equation}
L_{\rm brems} \sim 5.5 \times 10^{25} T^{1/2} R \tau^2 \; {\rm erg/s}
\end{equation}
Adopting the electron temperature of 0.86~keV obtained by the fit, 
the radius of the 
emission region is  $\sim 4.9 \times 10^{11}/\tau^{2}$ cm. For
an optical depth associated to the bremsstrahlung not greater than 1, 
the resulting radius of the emitting region is larger than the radius of the 
Roche lobe containing the neutron star. Therefore, in this scenario the 
black body component could be emitted by optically thick plasma at the 
magnetosphere, and the bremsstrahlung component could be produced by 
plasma of the stellar wind from the companion star, ionized by the X-ray 
emission from the pulsar.

It is also possible to model the soft excess with a Comptonization 
model. The corresponding luminosity of this component is 
$5 \times 10^{37}$ ergs/s. Knowing the temperature
of the seed photons and that of the hot electrons, the luminosity of the
emitting region and its optical depth,
it would be possible to infer the characteristic radius of the seed photons
emitting region, where the seed photons originate.
The COMPST model does not give the possibility to obtain the temperature
of the seed photons from the fit, therefore we derived  
this temperature assuming that it is the temperature at the inner radius of 
the accretion disk (assuming the standard model of Shakura \& Sunyaev 1973), 
under the hypothesis that the disk is truncated at the magnetospheric radius. 
Using the value of the magnetic field corresponding to the cyclotron line,
we obtain a temperature of the seed photons of $1.8 \times 10^{-2}$~keV.
From this we obtain that the radius of the region emitting
the seed photons is $36 \times 10^{8}$ cm. 
This value is in agreement within a factor 5
with that obtained from equation~4 where we have adopted $\mu_{30} = 11$, 
as deduced by the fit of the cyclotron feature. 
However, note this model might be unadequate to describe the whole emission 
mechanism, since at this cold temperatures and large optical depths, 
free-free absorption, that is not included in the COMPST model, could give a 
not negligible contribution.

An iron emission line is detected at about 6.5~keV. 
we find that it can be fitted by two gaussians at energies 6.1 and 6.6~keV, 
respectively. The gaussian at lower energy has
a large sigma ($\sim 0.95$~keV), whereas the gaussian at higher energies 
is much narrower.  We note that the width of the broad line component
cannot be explained by the Doppler effect induced by the fast rotation
at the magnetospheric radius. In fact considering a magnetic moment of
$1.1 \times 10^{31}$ Gauss cm$^3$ (as estimated by the cyclotron line
energy) we obtain $\Delta \nu \sim 0.1$~keV at 6.4~keV, that is much 
smaller than the measured width. 
Complex iron line features are not unusual in high 
mass X-ray binaries (see e.g.  Audley {\it  et al.} 1996, 
Ebisawa {\it  et al.} 1996).  The energy values that we obtained are
consistent with the iron features expected at 6.4 and at 6.7~keV,
within the associated errors. The usual interpretation is that the line at
6.7~keV is produced by highly ionized iron and probably is emitted in 
a hot corona, while the line at 6.4~keV corresponds to neutral iron and
might come from the radiation reflected by the accretion disk.  In this
case we would expect the 6.7~keV line to be broad because of the Compton
smearing in the hot corona, and the 6.4~keV line to be narrow given that
the disk is truncated at a large radius by the presence of the
magnetosphere.
We observe that our data of LMC~X--4 indicate an opposite behaviour,
with a broad 6.4~keV line and a narrow 6.7~keV line. Probably this feature
can be explained by line blending, for the presence of several iron line 
components with energies between 6.4 and 7~keV, or by the presence of a
line with a narrow core but broad wings. Certainly other high resolution
observations in the iron line region are needed to address this question.

All the models reported above need an absorption edge at about 1.25~keV.
The nature of this feature is highly uncertain. The energy of the edge
is close enough to the K-edge of high ionized Ne or neutral Mg, or to an 
L-edge of moderately ionized Fe.
If we attribute this feature to FeXVII (edge at 1.266~keV), 
the optical depth corresponding to this feature is:
\begin{equation}
\tau = \sigma \int N dx = 1 - 
          \frac{\phi_{\rm with-edge}}{\phi_{\rm no-edge}} \sim 0.1
\end{equation}
where $\phi_{\rm with-edge}$ and $\phi_{\rm no-edge}$ are the flux with and 
without the absorption edge, respectively, calculated in the 1.25-3.0~keV 
range. Assuming that the cross section of this process is comparable with 
that of the ejection of electrons from the K-shell 
($\sigma \sim 1.7 \times 10^{-16}$ cm$^2$) the corresponding 
iron column is $\sim 0.6 \times 10^{15}$ cm$^{-2}$.
This value could be compared with that deduced from the hydrogen column
obtained by the fit of the photoelectric absorption. In that case the hydrogen
column is $\sim 5 \times 10^{20}$ cm$^{-2}$ so the deduced iron column 
(assuming solar abundances) could be of the order of 
$5 \times 10^{15}$ cm$^{-2}$. The good agreement between
the two values indicates that the edge at 1.252~keV could be a real
feature produced by the iron around the neutron star. 
Other authors have also reported similar features. 
Parmar {\it  et al.} (2000) have found an absorption edge in the spectrum
of the low mass X-ray binary X~1822--371 at about 1.33~keV and optical depth
0.26. Moreover Iaria {\it  et al.} (2000) have observed the edge in the 
spectrum of the source 4U~1254--69 at 1.26~keV and optical depth 0.15.

\acknowledgments
This work was supported by the Italian Space Agency (ASI), by the Ministero
della Ricerca Scientifica e Tecnologica (MURST).

\clearpage

\begin{table}[h]
\begin{center}
\caption{Fit of LMC~X--4 spectrum during the high state in the energy 
band 0.12-100~keV.
Uncertainties are at 90\% confidence level for a single parameter.
$\tau_{\rm edge}$ is the absorption depth of the edge at the threshold.
kT$_{\rm BB}$ is the temperature of the black body. The normalization of the 
black body, N$_{\rm BB}$ is in units of L$_{36}$/D$^2_{50}$ where L$_{36}$ 
is the luminosity of the source in units of $10^{36}$ ergs/s and D$_{50}$ is 
the distance of the source in units of 50 kpc. 
kT$_{\rm Brems}$ is the temperature of the plasma emitting for
bremsstrahlung. The normalization of the bremsstrahlung, N$_{\rm Brems}$, 
is in units of $(1.27 \times 10^{-59}/(4 \pi D^2_{50})) \int N_e N_I dV $, 
where $N_e$ and $N_I$ are the electron and ion densities (cm$^{-3}$), 
respectively.
kT$_{\rm compst}$ is the temperature of the hot electrons Comptonizing cool 
photons. The COMPST normalization is in XSPEC (v.10) units. 
The power law normalization
is in units of photons~keV$^{-1}$ cm$^{-2}$ s$^{-1}$ at 1~keV. 
The (emission lines) gaussian normalization is in units 
photons cm$^{-2}$ s$^{-1}$.} 
\label{table1}
\vspace{0.5cm}
\small
\begin{tabular}{l|c|c|c|c} \hline \hline 
Parameters                        & model 1 & model 2 & model 3 & model 4 \\ \hline
$N_H (\times 10^{20} \; cm^{-2})$ & $5.0_{-0.4}^{+0.5}$
                                   & $5.0 \pm 0.5$   
                                     & $5 \pm 1$
                                       & $ 5 \pm 1$\\
                                 & & & & \\
Edge Energy (keV)                & $1.28 \pm 0.04$
                                   & $1.28_{-0.03}^{+0.02}$
                                     & $1.25 \pm 0.04$
                                       & $1.25 \pm 0.04$\\
$\tau_{\rm edge}$               & $0.4 \pm 0.1$
                                   & $0.4 \pm 0.1$
                                     & $0.47_{-0.1}^{+0.05}$
                                       & $0.5 \pm 0.1$\\
                                 & & & & \\
kT$_{\rm BB}$ (keV)                  & -- 
                                   & -- 
                                     & $(6.3 \pm 1.5) \times 10^{-2}$
                                       & $(6.3 \pm 1.5) \times 10^{-2}$\\
N$_{\rm BB}$                        & -- 
                                   & -- 
                                     & $2.4_{-1.2}^{+4.0} \times 10^{-2}$
                                       & $2.4_{-1.2}^{+4.0} \times 10^{-2}$\\
kT$_{\rm Brems}$ (keV)               & -- 
                                   & -- 
                                     & $0.8_{-0.2}^{+0.3}$
                                       & $0.8 \pm 0.2$\\
N$_{\rm Brems}$                   & -- 
                                   & -- 
                                     & $4_{-1}^{+5} \times 10^{-2}$
                                       & $(4 \pm 1) \times 10^{-2}$\\
                                 & & & & \\
kT$_{\rm compst}$ (keV)              & $0.9_{-0.3}^{+4}$
                                   & $9_{-6}^{+70}$
                                     & --
                                       & --\\
$\tau_{\rm compst}$                          & $16 \pm 5$
                                   & $5_{-2}^{+6}$
                                     & --
                                       & --\\
N$_{\rm compst}$           & $2.1_{-0.2}^{+0.3} \times 10^{-2}$
                                   & $1.7 \pm 0.2$
                                     & --
                                       & --\\
                                 & & & & \\
Photon Index                     & $0.5_{-0.2}^{+0.1}$
                                   & $0.5 \pm 0.1$ 
                                     & $0.6 \pm 0.1$
                                       & $0.64_{-0.04}^{+0.03}$\\
N$_{\rm power\;law}$                    & $(1.1 \pm 0.2) \times 10^{-2}$
                                   & $7_{-1}^{+2} \times 10^{-3}$
                                     & $1.0_{-0.1}^{+0.2} \times 10^{-2}$
                                       & $9.9_{-0.8}^{+0.5} \times 10^{3}$\\ 
Smooth Width (keV)               & $9 \pm 4$
                                   & $17 \pm 3$
                                     & $9 \pm 4$
                                       & $15 \pm 3$\\ 
$E_{\rm cut}$ (keV)                  & $18 \pm 1$ 
                                   & $18 \pm 1$
                                     & $18 \pm 1$
                                       & $18.7 \pm 0.5$\\    
$E_{\rm fold}$ (keV)                 & $18 \pm 3$
                                   & $12.2 \pm 0.4$  
                                     & $17_{-3}^{+4}$
                                       & $12.7 \pm 0.3$\\
                                 & & & & \\
$E_{\rm Fe}$ (keV) 1st gaussian      & $6.1 \pm 0.3$
                                   & $6.1_{-0.3}^{+0.1}$
                                     & $6.2_{-0.8}^{+0.2}$
                                       & $6.2_{-0.7}^{+0.2}$\\ 
$\sigma_{\rm Fe}$ (keV)              & $0.9_{-0.2}^{+0.4}$
                                   & $1.1_{-0.2}^{+0.5}$
                                     & $0.8 \pm 0.4$
                                       & $0.9_{-0.5}^{+0.3}$\\
$I_{\rm Fe}$ (ph cm$^{-2}$ s$^{-1}$) & $9_{-4}^{+150} \times 10^{-4}$
                                   & $8_{-3}^{+5} \times 10^{-4}$
                                     & $6_{-4}^{+10} \times 10^{-4}$
                                       & $6_{-4}^{+3} \times 10^{-4}$\\
EW$_{\rm Fe}$ (eV)                   & 169 & 183 & 215 & 267 \\
$E_{\rm Fe}$ (keV) 2nd gaussian      & $6.6 \pm 0.1$
                                   & $6.6 \pm 0.1$
                                     & $6.6 \pm 0.1$
                                       & $6.6 \pm 0.1$\\ 
$\sigma_{\rm Fe}$ (keV)              & $(0.1 \pm 0.2) \times 10^{-2}$ 
                                   & $(0.1 \pm 0.1) \times 10^{-2}$
                                     & $7_{-7}^{+30} \times 10^{-2}$
                                       & $7_{-7}^{+25} \times 10^{-2}$\\
$I_{\rm Fe}$ (ph cm$^{-2}$ s$^{-1}$) & $3_{-1}^{+5} \times 10^{-4}$ 
                                   & $(2 \pm 1) \times 10^{-4}$
                                     & $2.3_{-0.4}^{+4} \times 10^{-4}$
                                       & $2_{-1}^{+3}\times 10^{-4}$\\
EW$_{\rm Fe}$ (eV)                   & 64 & 64 & 67 & 69\\
                                 & & & & \\
$E_{\rm cycl}$ (keV)                  & $100_{-20}^{+100}$ 
                                   & --
                                     & $100_{-15}^{+80}$
                                       & -- \\
$\sigma_{\rm cycl}$ (keV)             & $60_{-20}^{+100}$ 
                                   & --
                                     & $45_{-20 }^{+140}$
                                       & --\\
A$_{\rm cycl}$                        & $1.0_{-0.2}^{+0.4}$ 
                                   & --
                                     & $1.0_{-0.3}^{+0.4}$
                                       & --\\
                                 & & & & \\
$\chi^2$/d.o.f.                  & 614/536 
                                   & 636/539 
                                     & 614/535 
                                       & 638/538 \\ \hline
\end{tabular}
\end{center}
\end{table} 
\clearpage

\section*{FIGURE CAPTIONS}
\bigskip

\noindent
{\bf Figure 1:} Upper panel: LMC~X--4 light curve in the energy band 
1.5-10.5~keV band (MECS data). The bin time is 300 s.
Lower panel: Corresponding hardness ratio [4.5--10.5~keV]/[1.5--4.5~keV].
\\
{\bf Figure 2:} Upper Panel: Residuals in unit of $\sigma$ 
with respect to model 2 of Table 1
(without the cyclotron absorption line correction). Lower Panel: 
Residuals in unit of $\sigma$
with respect to model 1 of Table 1. In this case the cyclotron 
absorption line correction is taken into account.
Different symbols have been used to distinguish between HPGSPC data
(filled triangles) and PDS data (open triangles).
\\
{\bf Figure 3:} Pulse-averaged count spectrum (0.12-100~keV) of LMC~X--4 and 
residuals in units of $\sigma$
corresponding to model 3 of Table 1. In this case the soft part of 
the spectrum is fitted
with bremsstrahlung plus black body. The cyclotron absorption 
line is also included. 
\\
{\bf Figure 4:} Unfolded energy spectrum and best fit model 
corresponding to model 3 of Table 1. The solid line with the data on top
is the total spectrum, the dashed lines are the blackbody (left) and the
bremsstrahlung component (right) respectively,
the dot-dot-dot-dashed line is the power-law component, 
the dot-dashed line is the gaussian at 6.1~keV, and the dotted line is the 
gaussian at 6.6~keV.
\\
{\bf Figure 5:} Pulse-averaged count spectrum (0.12-100~keV) of LMC~X--4 
and residuals in units of $\sigma$
corresponding to model 1 of Table 1. In this case the soft part of 
the spectrum is fitted
with a Comptonization. The cyclotron absorption line is included.
\\
{\bf Figure 6:} Unfolded energy spectrum and best fit model
corresponding to model 1 of Table 1. The solid line with the data on top
is the total spectrum, the dashed line is the comptonized emission component,
the dot-dot-dot-dashed line is the power-law component, 
the dot-dashed line is the gaussian
at 6.1~keV, and the dotted line is the gaussian at 6.6~keV.

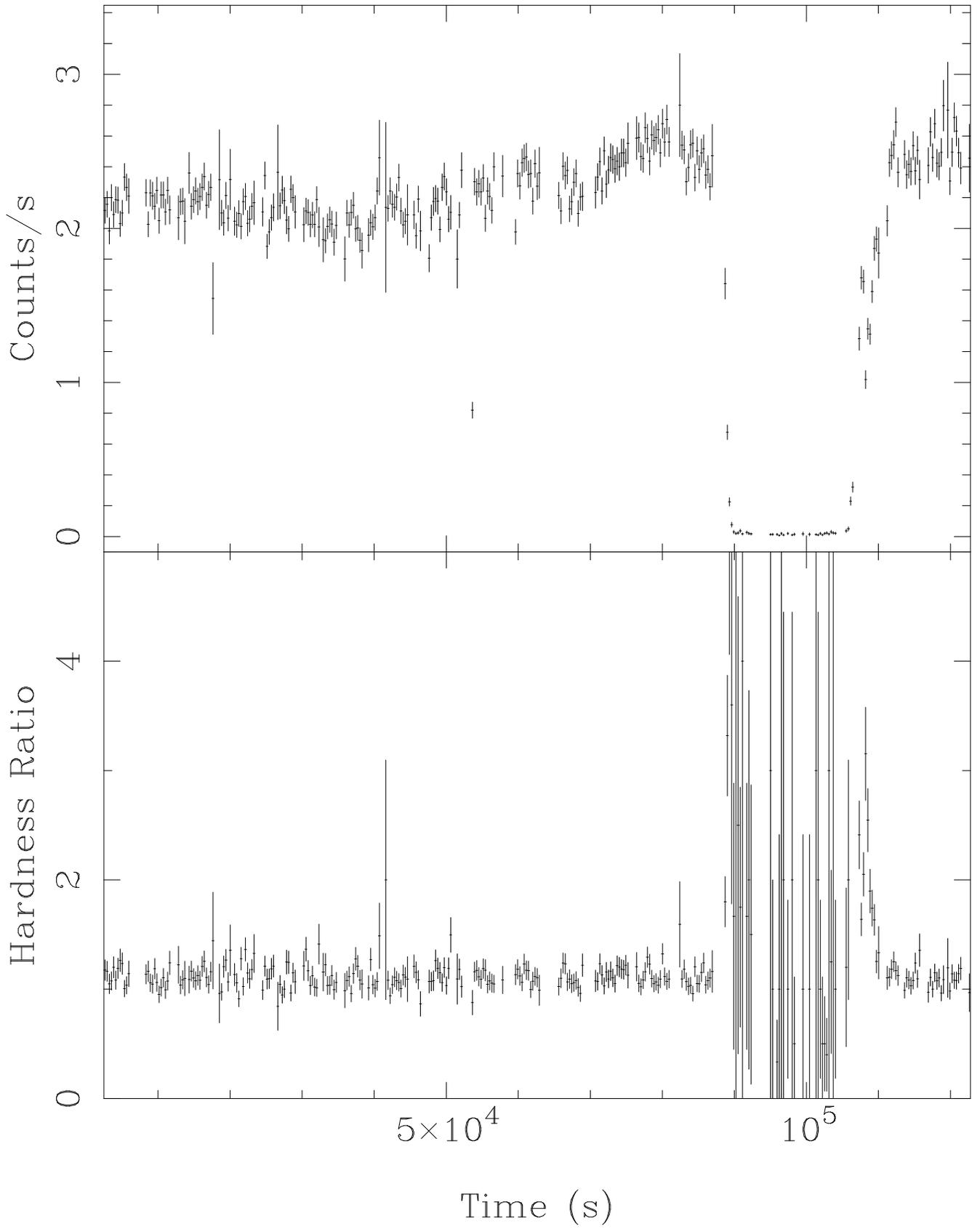
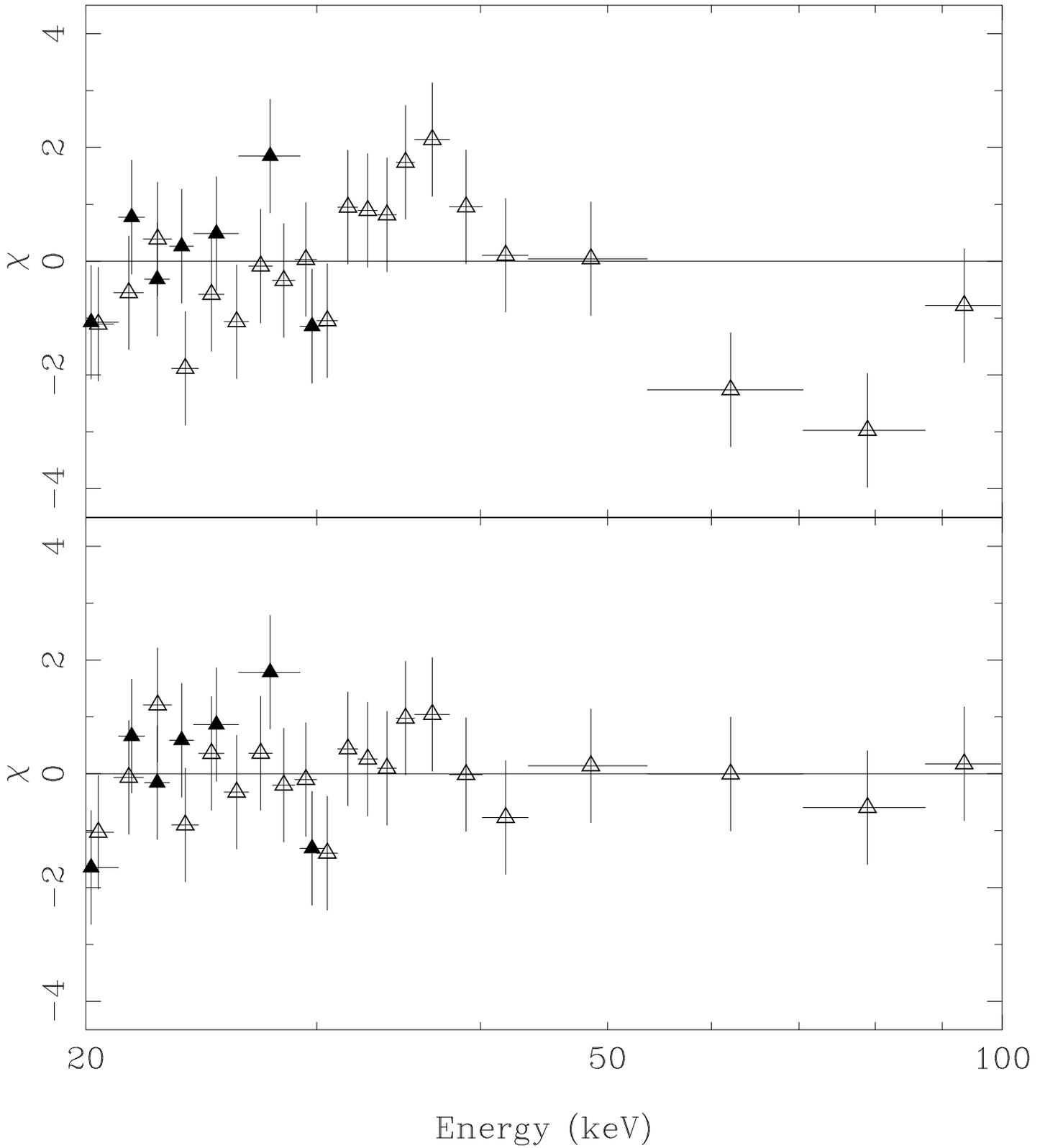
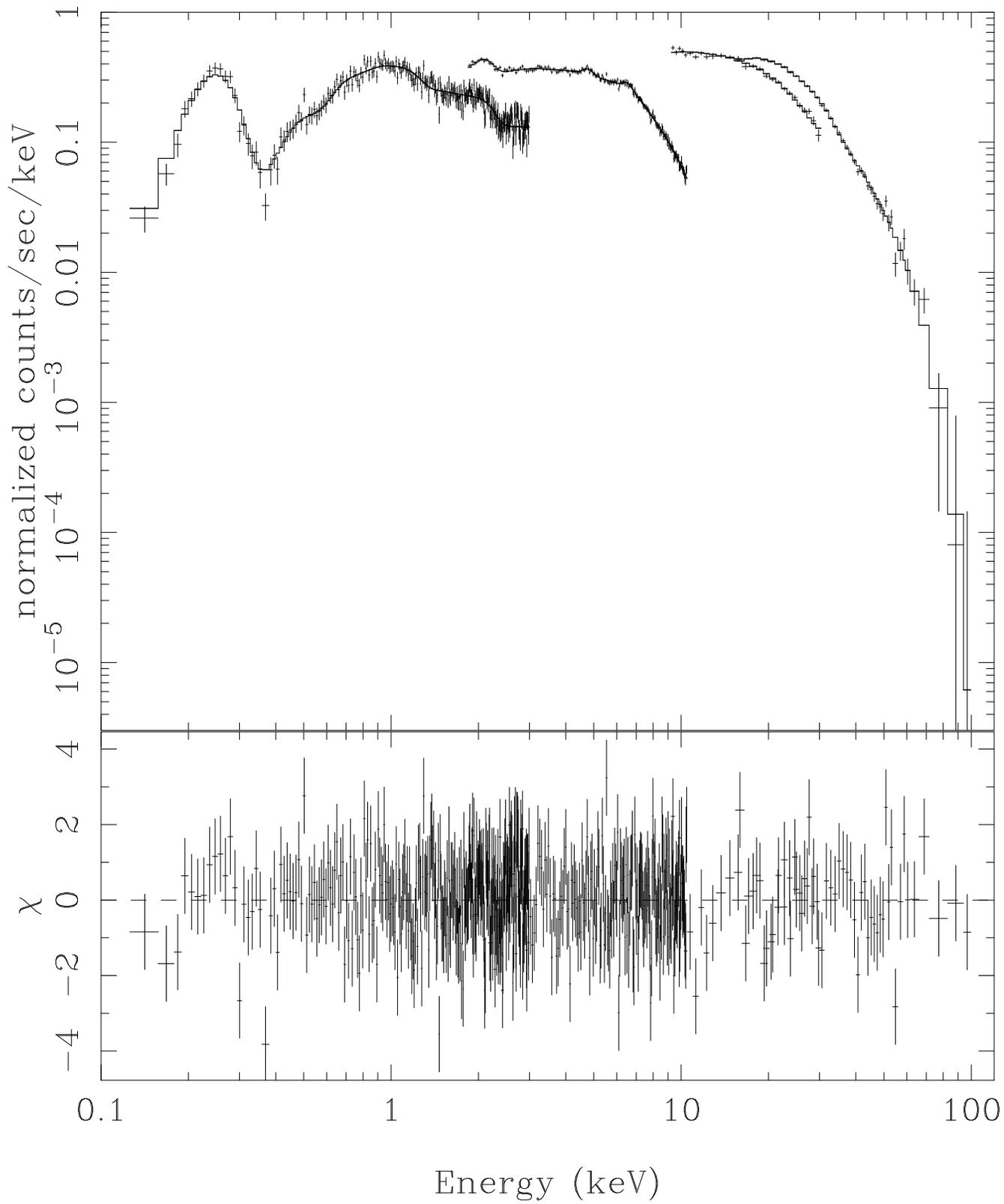
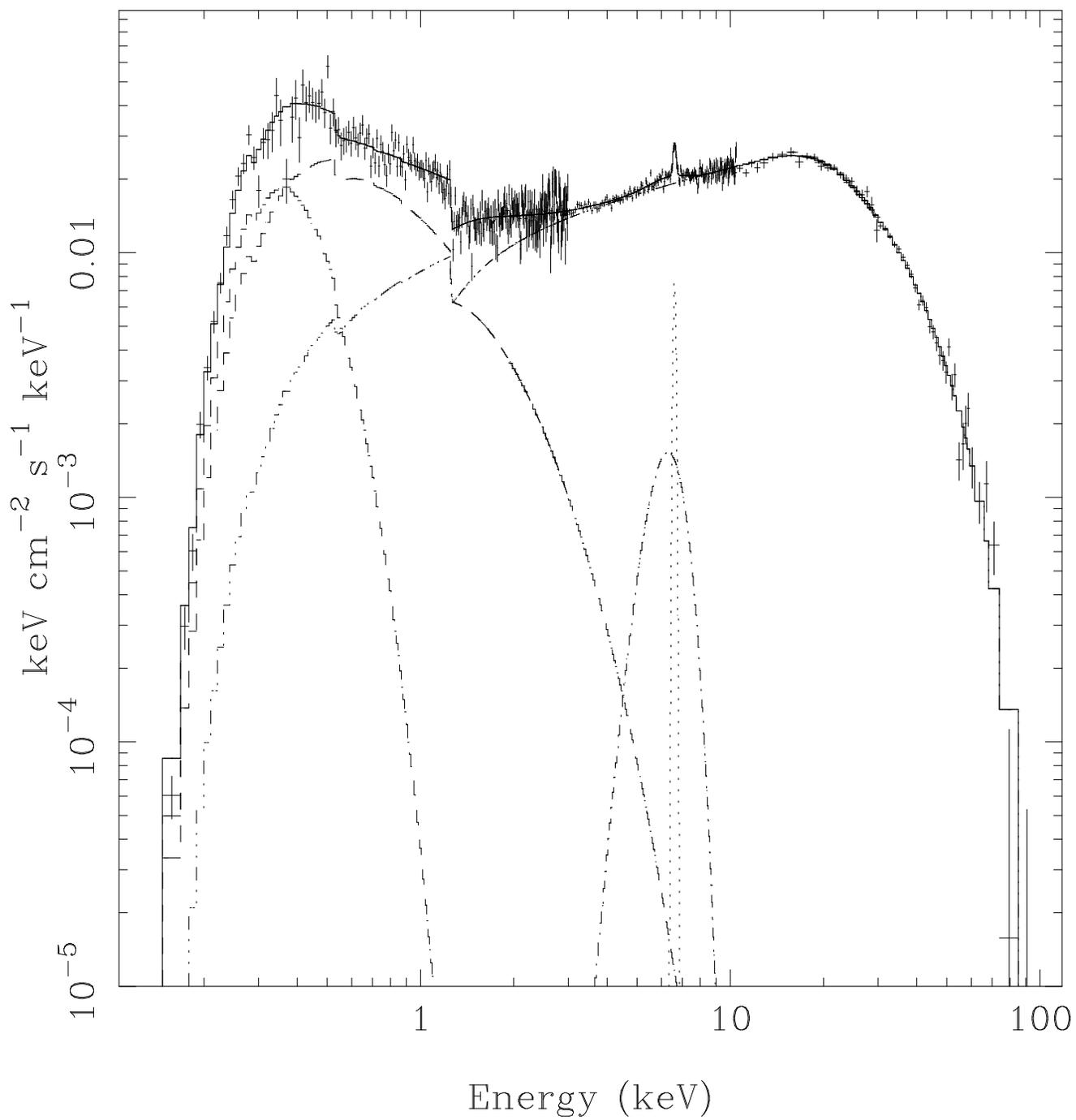
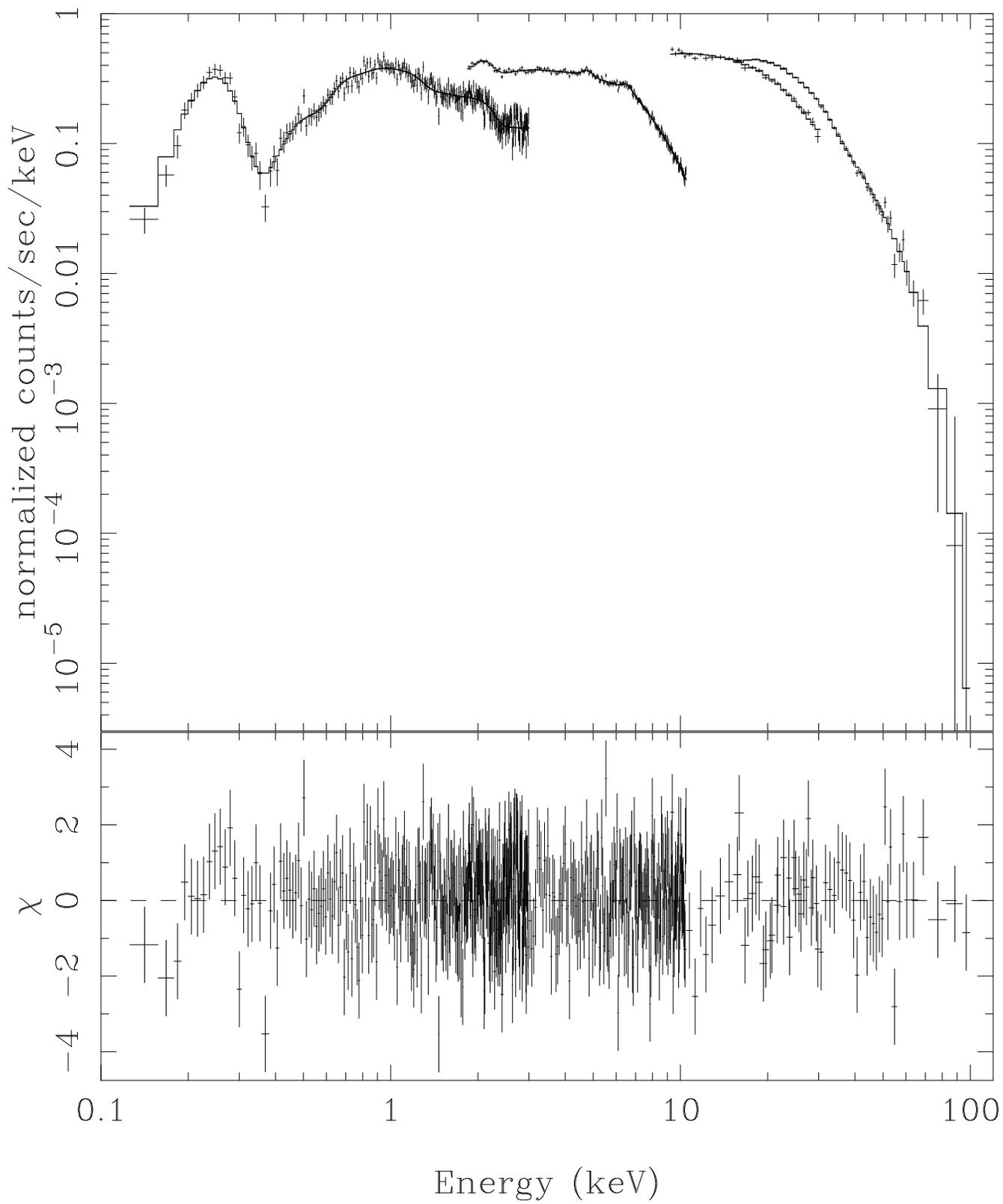
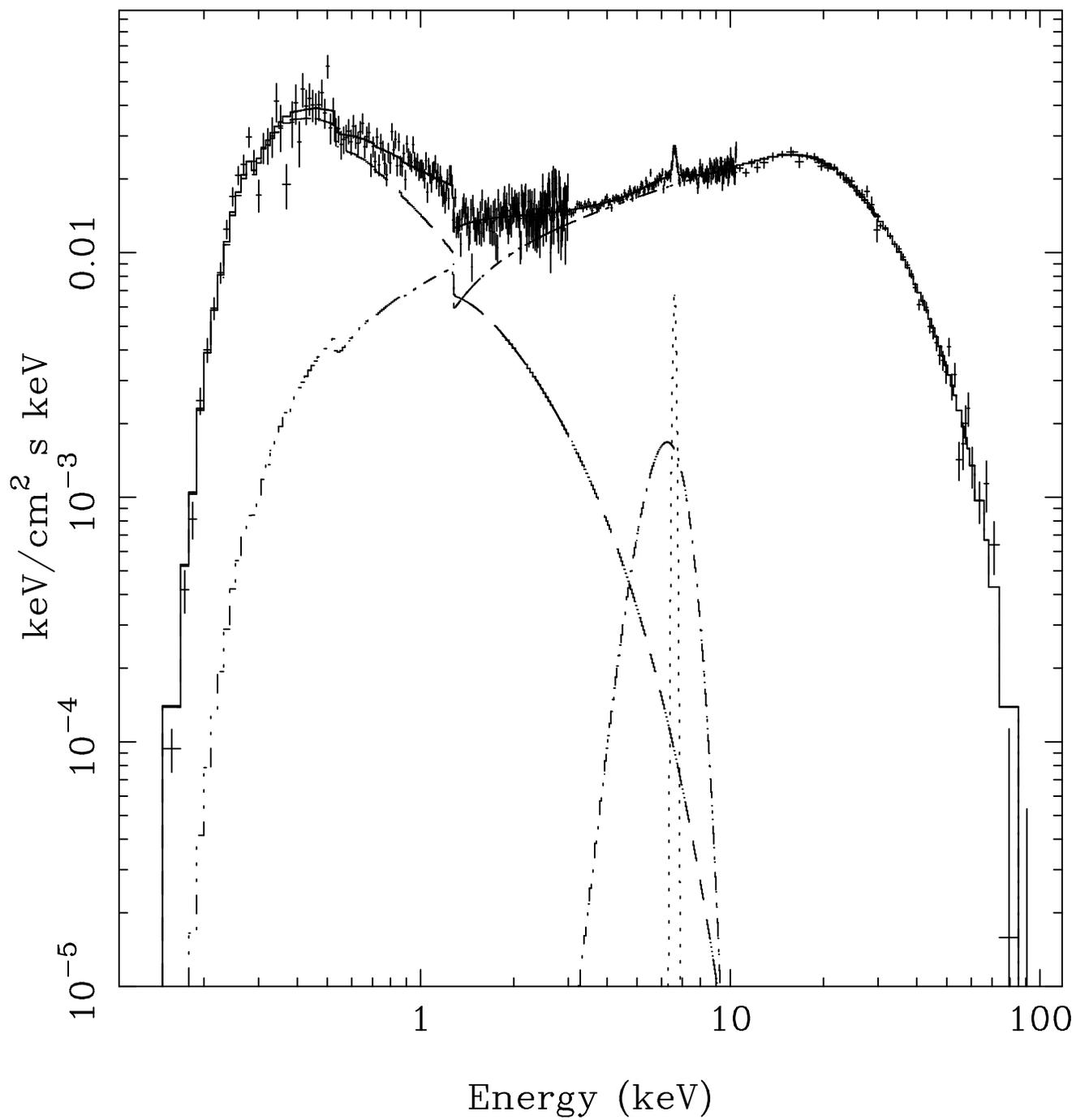
\end{document}